\documentclass[12pt]{iopart}
\usepackage{iopams,times,graphicx,colordvi,t1enc,amssymb,mathptmx}

\begin{document}

\title[Orbital and spin magnetization of quantum dots and rings.]
      {Orbital and spin magnetization of a confined electronic
       system in the transition between a quantum dot and a ring.}
\author{Gabriel Vasile, Vidar Gudmundsson, and Andrei Manolescu}
\address{Science Institute, University of Iceland,
             Dunhaga 3, IS-107 Reykjavik, Iceland}
%
%

\begin{abstract}

In order to understand the orbital and spin magnetization of a confined
electronic system we analyze these ground state properties in the
transition from a quantum dot to a quantum ring of finite thickness. The
Coulomb interaction between the electrons is treated in the Hartree
and Hartree-Fock approximations and special care is taken to include
also the contributions of the nonlocal current to the summation of the
magnetic moments of the occupied states.  We identify changes in the
many-body structure of the ground state and in the magnetization curves
versus the magnetic field and other parameters characterizing the system.
We compare the results of two models for quantum dots (or rings), one
with the electrons moving continuously in the system, and one with the
electrons moving on a lattice.

\end{abstract}
\section{Introduction}
Recently the magnetization of an electronic systems 
consisting of few hundred electrons confined in each dot of an array
of quantum dots has been measured \cite{Swhwarz02:6875}. Researchers
hope that the magnetization measurements will turn out to be an additional 
method to probe the ground state of small electronic systems.
The temperature scales of different types of magnetization oscillations 
in circularly symmetric or slightly asymmetric quantum dots with many 
noninteracting electrons has been modeled \cite{Bogachek01:115323}.
The relation of the magnetization and the persistent currents in quantum 
dots and rings with many noninteracting electrons has been investigated 
in a model by Tan and Inkson \cite{Tan99:5626}, 
and by Aldea et al.\ \cite{Aldea02:xx}. The effects of an
impurity on the magnetization of two interacting electrons in a square
dot with hard walls have been reported \cite{Sheng98:152}.
The effects of the geometry of quantum dots on the magnetization 
for noninteracting or electrons interacting in the Hartree Approximation (HA) 
have been reported by Magnúsdóttir and Gudmundsson \cite{Magnusdottir00:10229}.

For few electrons, the exchange interaction and their spin configuration
is important when considering the influence of the Coulomb interaction 
on the magnetization. Here we use the Hartree-Fock Approximation (HFA)
to investigate the magnetization of confined systems with 4 - 6 electrons 
in the transition from dots to rings in order to understand better the
connection between features in the magnetization and properties of the
ground state in rather pure systems that will become the
subject of experiments in close future.   

\section{Models}
We consider a quantum dot with few electrons confined by the potential
\begin{equation}
      V_{\mbox{conf}}(r,\phi )=\frac{1}{2}m^*\omega_0^2r^2\left[
      1+\sum_{p=1}^{p_{max}}\alpha_p\cos (2p\phi )\right]
      +V_0\exp(-\gamma r^2),
\label{V_conf}
\end{equation}
where the parameters $\alpha_p$ can be used to break the circular
symmetry of the parabolic confinement, and $\gamma$ and $V_0$ control
a circular hill in the center of the dot in order to change it into a ring.
The electron-electron interaction is treated in the Hartree-Fock 
Approximation (HFA) at a finite temperature $T$.  We only consider 
the $z$-spin component of each electron. The effective
single-electron Schrödinger equation is solved iteratively in the mathematical
basis of the Fock-Darwin wave functions \cite{Fock28:446,Magnusdottir00:10229}.  
In the HFA we have a nonlocal equation of motion and no explicit 
single-electron Hamiltonian. Thus, the current density does not have 
the same simple local expression as in the Hartree approximation.
Instead, by defining the current density as
\begin{equation}
      {\mathbf j}=-e\dot{\mathbf r}=\frac{ie}{\hbar}\left[ H,{\mathbf r}\right]
\end{equation}
we construct the matrix element of the contribution 
of each Hartree-Fock state $|\alpha )$ of the orbital magnetization operator
${\mathbf M}_o={\mathbf r}\times{\mathbf j}$ and sum up  
the total magnetization of the system
\begin{equation}
      {\mathbf{\cal M}}_o=\sum_{\alpha}f_{\alpha}(\alpha |{\mathbf M}_o|\alpha ), 
\end{equation}
where $f_{\alpha}$ is the occupation of the state $|\alpha )$ according to the 
equilibrium Fermi distribution.  With this model we cannot
affort to describe more than a few electrons in a noncircular
quantum dot. The larger the deviation is from the parabolic
confinement with circular symmetry, the bigger the basis
has to be we use for representing the eigenstates in.

\bigskip

In order to see what can happen in a system with more electrons,
and also to verify the main features in our calculation, 
we compute the magnetization  for a tight-binding
lattice model with 100 sites with coordinates $(x,y) \equiv (na,ma)$, with 
$n,m=1,2,...,10$, $a$ being the lattice constant \cite{Aldea02:xx}.  
In this model the Hamiltonian is written as
\begin{eqnarray}
      \hspace{-14mm}H&&\hspace{-10mm}=\sum_{n,m} \left[
      V_{nm} \vert n,m \rangle\langle n,m\vert
      + t \left(e^{i\pi m\phi}\vert n,m\rangle\langle n+1,m\vert
      + e^{-i\pi n\phi}\vert n,m\rangle\langle n,m+1\vert + h.c. \right) \right] \nonumber\\
      &&\hspace{-10mm}+U_c \sum_{n,m\neq n',m'} {\nu_{n'm'}\over {\sqrt{(n-n')
      ^2+(m-m')^2}}} \,\, \vert n,m\rangle \langle n',m'\vert \,, 
\end{eqnarray}
where $V_{nm}$ is a potential energy on each site, $t$ is the energy of hopping 
between nearest-neighbor sites, $\phi$ is the magnetic flux through
the unit cell in units of magnetic flux quanta, $U_c$ is the 
Coulomb energy in units of $t$, and $\nu_{n'm'}$ is the site occupation.
The inclusion of the Coulomb energy corresponds here to the Hartree 
approximation.  The tight binding model is indeed more primitive,
but with the number of sites chosen here  it can reasonably describe 
a quantum dot of 100 nm width in a magnetic field up to 5 T. 
More details on the applicability
of the tight-binding model can be fond in \cite{Aldea02:xx} and 
references therein.  The main advantages of this model is that we can
easily use it for more electrons than the continuous model
with the Darwin-Fock basis.
Also, we can easily define potential barriers inside, through the on-site
potentials $V_{nm}$, and give various shapes to the dot, or change
it into a ring.  

\section{Results}
\begin{figure}[htq]
   \begin{center}
      \includegraphics[width=15.5cm]{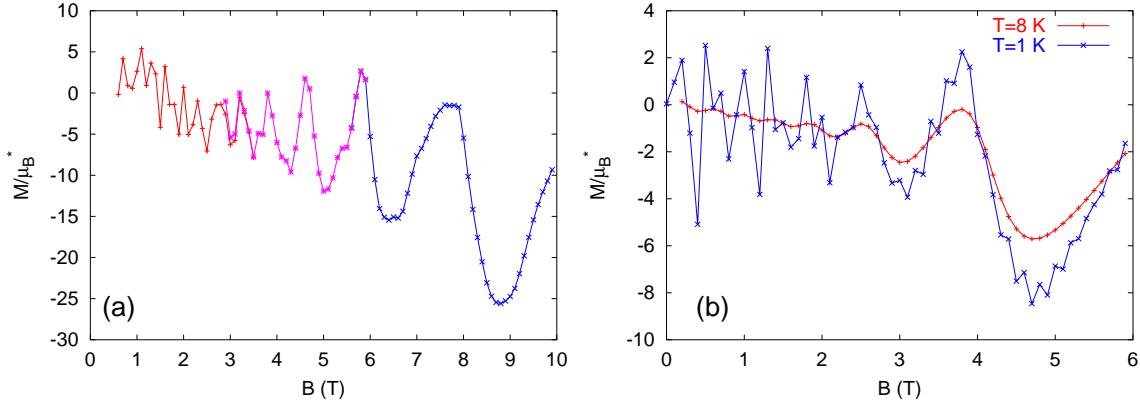}
   \end{center}
\caption{The orbital magnetization of 230 electrons at $T=1$ K (a), and 40 electrons 
         at $T=1$ and 8 K (b) in a quantum dot
         in the Hartree approximation in units of $\mu^*_B$. 
         Due to the high number of electrons
         and large $B$ range for $N=230$ three different sizes of the basis sets
         are used in the calculation and indicated with color on the graph. 
         $\hbar\omega_0=6$ meV for $N=230$ and $\hbar\omega_0=3.37$ meV for $N=40$, 
         no spin, and circular symmetry.}
\label{fig_230}
\end{figure}

In accordance with experiments on magnetization of quantum dots \cite{Swhwarz02:6875}
we use GaAs parameters, $m^*=0.067m_e$, $\kappa=12.4$, and $g^*=-0.44$.
In Fig.\ \ref{fig_230} we show the magnetization for 230 and 40 
electrons in a quantum dot, in the HA.  For that many electrons
we do not expect important exchange effects.
For magnetic fields above 2 T we see de Haas-van Alphen oscillations and for
the 40 electrons at $T=1$ K sharper oscillations are superimposed due to the Aharonov-Bohm 
effect and finer details in the density of states \cite{Bogachek01:115323}.  
Clearly, the experimental situation is a bit more complicated \cite{Swhwarz02:6875},
especially when it comes to comparing the magnetization in the regimes of high and low
magnetic field.   

\begin{figure}[htq]
   \begin{center}
      \includegraphics[width=15.5cm]{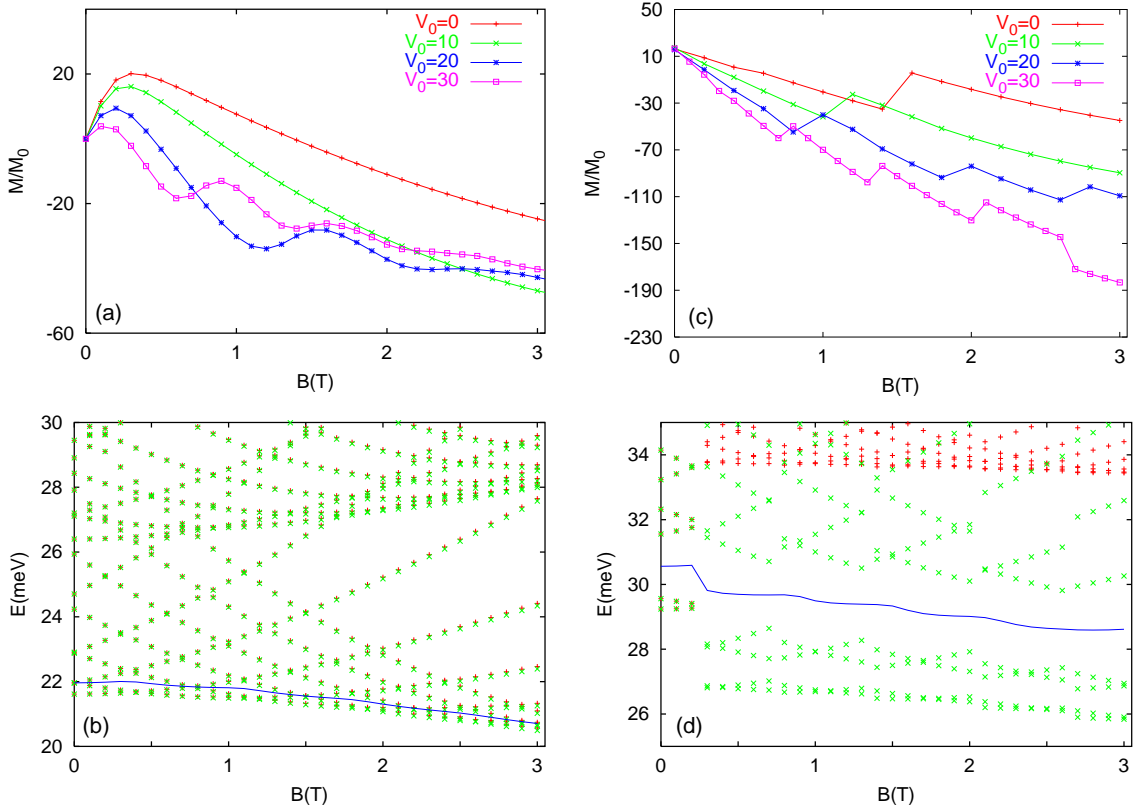}
   \end{center}
\caption{For the noninteracting system the magnetization in units of $M_0=\mu_B$ 
         (a), and the energy
         spectrum (b), and for the interacting system the magnetization (c)
         and the effective energy spectrum (d). The interaction is treated
         in the Hartree-Fock approximation, and the system evolves from a dot
         to a ring with increasing $V_0$.  The chemical potential is
         indicated with a continuous curve in the subfigures for the energy
         which are for $V_0=30$ meV.  
         $\hbar\omega_0=3.37$ meV, and $N=4$. $T=1$ K.}
\label{figGV}
\end{figure}

For only few electrons in a dot their spin configuration becomes essential 
and the exchange interaction between them can not be neglected. To better appreciate
the effects of the interaction we display in Fig.\ \ref{figGV}  the magnetization of
a quantum dot with 4 electrons as it is changed into a quantum ring by increasing
$V_0$, the height of the central hill in the confinement potential
(eq.\ \ref{V_conf}).
The energy spectrum for a infinitely narrow ring is periodic in the magnetic
flux through it. But in the case of a ring of finite width the energy spectrum
for a single electron is a mixture of the ring spectrum and the Fock-Darwin 
spectrum for a parabolic circular quantum dot, see Fig.\ \ref{figGV}b. For 
low magnetic field the spectrum is almost periodic for the lowest energy 
levels, and at a higher field a Landau band structure emerges as the effective
magnetic length becomes comparable to or smaller than the width of the ring.   
The magnetization in Fig.\ \ref{figGV}a starts to show oscillations as the
system becomes more ring like, but never becomes exactly periodic as is known
for a thin ring.

The electron-electron interaction changes the magnetization considerably, especially the
exchange interaction. If we first look at the pure dot confinement,
$V_0=0$ in Fig.\ \ref{figGV}c, we notice a jump in ${\cal M}_o$ between 1
and 2 T, but the curve in the Hartree approximation, not shown here,
is smooth. In the HA, at finite $T$, when the magnetic field increases
the electrons are gradually pushed to occupy states 
with a higher angular momentum quantum number $M$. In the HFA, 
at low $T$, the states are either occupied or empty, and the sharp 
differences in the exchange energy lead to jumps in the magnetization as 
$B$ increases. 

In a quantum ring, i.\ e.\ for large values of $V_0$, the electrons are barred
from the center region so states with low $M$ are not occupied for low $B$.
When the magnetic field is increased more complex reorganization of the occupation
of the $M$ states occurs and thus more jumps are seen in the orbital magnetization
${\cal M}_o$. The effective single-electron Hartree-Fock energy spectrum shown in
Fig.\ \ref{figGV}d reflects this fact. For all but the lowest magnetic field
values we have a strong enhancement of the spin splitting, the system is spin
polarized, and we have a large gap around the chemical potential $\mu$.    

For the lowest two values of the magnetic field we do not get the system
spin-polarized and we have not been able to attain the correct ground state
as can be confirmed by the fact that we do not get a vanishing orbital 
magnetization for $B\approx 0$. Here the $\pm M$ states are not symmetrically occupied
as should be. This is a common problem with the HFA, the system gets trapped
in a final state that is not the ground state and we have to start the iterations
with many different initial conditions to map out possible final states in the
calculation. It should though be mentioned here that all the final states here
have the same angular symmetry as the initial state, we shall come back to this
point below. 

\bigskip

\begin{figure}[htq]
   \begin{center}
      \includegraphics[width=15.5cm]{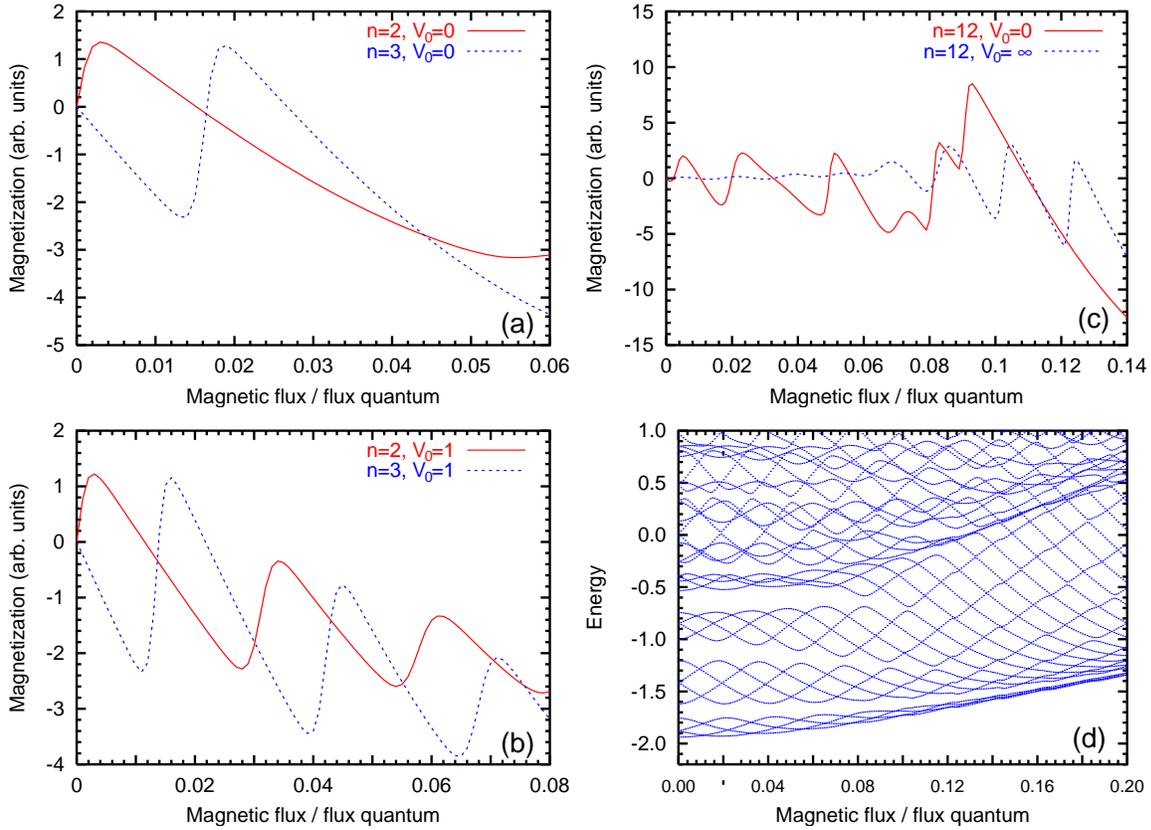}
   \end{center}
\caption{The magnetization of a square shaped quantum dot in the Hartree
         approximation for 2 and 3 electrons (a), square shaped quantum ring
         with 2 and 3 electrons (b), and 12 electrons in a dot or a ring (c).
         The energy spectrum corresponding to the ring with 12 electrons (d).}
\label{figA}
\end{figure}

In Fig.\ \ref{figA} we show some results obtained with the
tight-binding model for a square quantum dot with $n$=2, 3, 
and 12 electrons.
Since in the tight-binding model the electrons have no spin degree of 
freedom the magnetization for
2 electrons in Fig.\ \ref{figA}a corresponds to the results for 4 electrons 
in Fig.\ \ref{figGV}a except for the influence of the exchange. (Direct correspondence
can be found to the results of Magnúsdóttir \cite{Magnusdottir00:10229}).
By putting an energy $V_{nm}=V_0$ to 16 central sites we define
a square quantum ring.  The values of $V_0$ are shown in Fig.\ \ref{figA} in units
of the hopping energy $t$.  Fig.\ \ref{figA}b, reproduces the oscillations
seen in Fig.\ \ref{figGV}a for the noninteracting electrons in a ring, despite,
the differences in the geometrical shape, but, as expected, it does not show the
sharp jumps created by the exchange interaction seen in Fig.\ \ref{figGV}c.   
In the case of 12 electrons in a square dot ($V_0=0$) 
the magnetization displayed in Fig.\ \ref{figA}c
shows the first signs of de Haas-van Alphen oscillations caused by the density of
states at the Fermi level. Interestingly, the oscillations are subdued in the
corresponding ring except in the regime corresponding to the last oscillation
in ${\cal M}_o$ for the dot, when the chemical potential is in the lowest band in the
energy spectrum seen in Fig.\ \ref{figA}d.  

\bigskip

\begin{figure}[htq]
   \begin{center}
      \includegraphics[width=15.5cm]{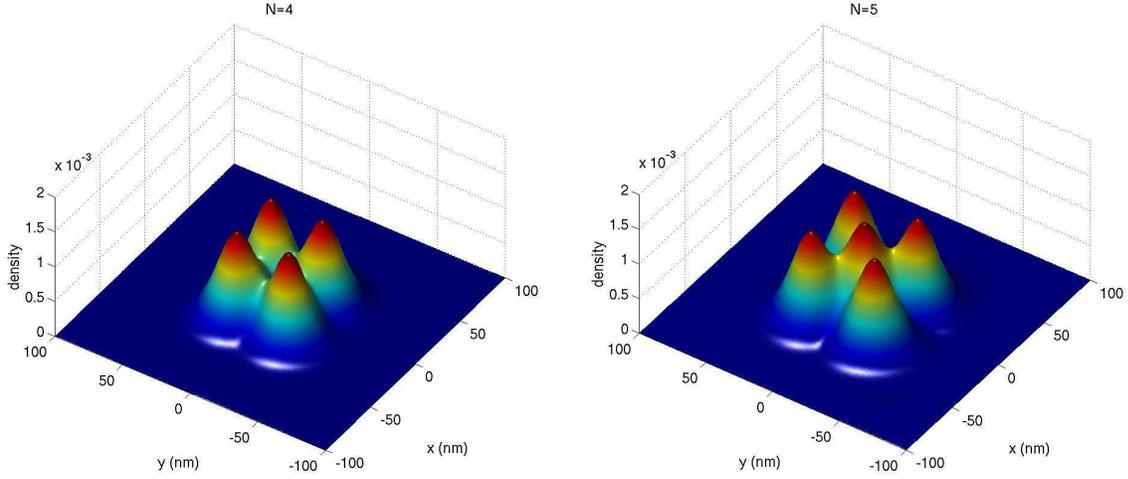}
   \end{center}
\caption{The electron density for a quantum dot with 4 or 5 
         spin polarized electrons and a slight square deviation,
         $\alpha_1=0$, $\alpha_2=0.1$. $T=1$ K, $\omega_0=3.37$ meV,
         $g^*=-0.44$.}
\label{fig4-5}
\end{figure}

\begin{figure}[htq]
   \begin{center}
      \includegraphics[width=15.5cm]{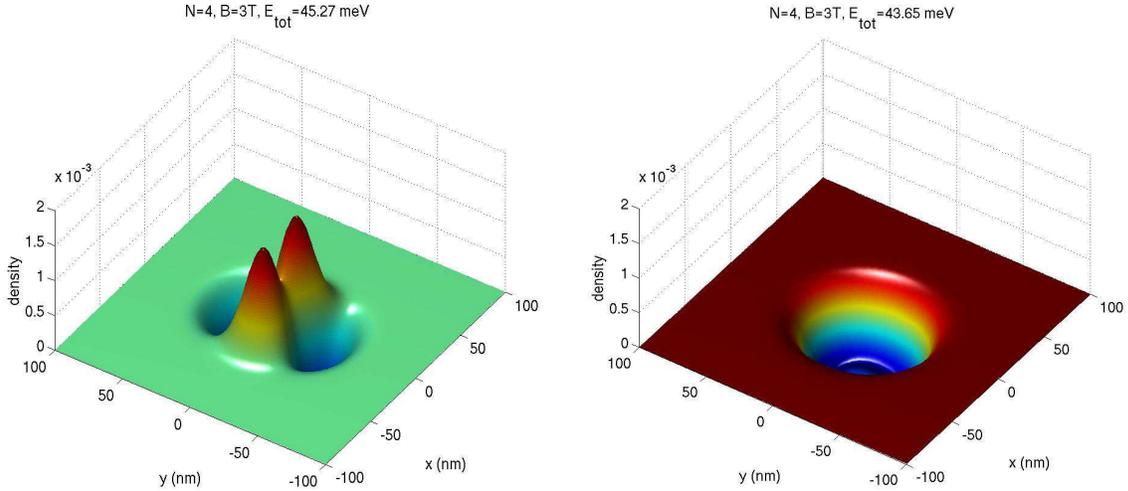}
   \end{center}
\caption{The spin-density for an excited state and the ground state of
         a quantum dot with a slight elliptical deviation, $\alpha_1=0.1$,
         $\alpha_2=0$. $T=1$ K, $\omega_0=3.37$ meV, $g^*=-0.44$.}
\label{figspin}
\end{figure}

In GaAs the spin magnetization is generally much smaller than the orbital one
due to the small $g$ factor and the small effective mass, but in contrast to the
orbital magnetization the fine structure of the spin magnetization can be quite 
sensitive to the geometrical shape of the system.  In Fig.\ \ref{fig4-5} we 
show the electron density for 4 and 5 spin polarized electrons, for a dot with 
slight square deviation in the confinement potential, 
calculated again using the continuous model. 
These are the ground states and we can correlate the number of peaks with
the number of electrons, though that is generally not unique. In the case
of 4 spin polarized electrons in a slightly elliptic confinement we have a 
{\lq\lq}homogeneous{\rq\rq} spin density seen in Fig.\ \ref{figspin}.
In the same dot we also find an excited state with no total spin polarization 
but a nice internal spin-density wave shown in Fig.\ \ref{figspin}. 
It should be added here that we have repeated the calculations for different
shapes, number of electrons and initial parameters. Each time we start with
a circular symmetric system we get an end state of the same geometric symmetry.
We assume this stability can be explained by the use of polar coordinates 
and circular basis functions. As soon as the initial state is slightly 
perturbed away from the circular symmetry the interaction can cause large
changes in the symmetry of the end state.

\section{Conclusions}

The magnetization is sensitive to the shape of the dot, to the electron-electron
interaction, and to the number of 
electrons.  The deviation from the circular symmetry involves more $M$ states
which for few electrons gives the potentiality for more 
reconfigurations thus producing more oscillations in the magnetization
as a function of $B$. The effect is stronger in a ring.
The Hartree interaction tends to expand the density and to change the
shape, by rounding the corners when the deviation from circular symmetry is
slight, or by stretching it in dots with large deviation in order
to lower the electrostatic energy. 
For few electrons the exchange interaction produces abrupt jumps of the 
magnetization as the exchange energy is very dependent on the exact
occupation of the single-electron states.

The oscillations of the magnetization, at low $T$, are determined by 
discontinuities of the occupation of the single-particle states,
i.\ e.\ by the jumps of $E_F$ or $\mu$.  Such oscillations become the 
Shubnikov-de Haas
oscillations at strong magnetic field.  In addition, smaller oscillations
occur due to the changes below $E_F$, in the occupied states. 
These oscillations can be understood in several ways: the magnetization
of an individual state follows the slope of the energy vs.\ $B$, 
which is related to the chirality \cite{Aldea02:xx}; 
changes of the average angular-momentum quantum number $M$ 
with $B$; the shape of the orbits, related to changes in the wave functions
with the magnetic field $B$.

We are aware of the fact that correlation effects may have their influence
on the magnetization of dots, as they have for the two-dimensional electron
system \cite{Meinel01:121306},
both affecting the finer details and the exact location
of reconfigurations in the ground state, but as experimental methods have to
be refined quite a bit to observe finer details in dots with few electrons we
have limited our handling of the electron-electron interaction here to the 
direct and exchange effects.   

\ack
We acknowledge instructive discussions with A.\ Aldea and M.\ Schwarz.
The research was partly funded by the Icelandic Natural Science Foundation,
and the University of Iceland Research Fund.

\section*{References}
\bibliographystyle{prsty}
\bibliography{mod_qd.bib}

\end{document}